\documentclass[12pt]{article}
\usepackage{amsmath,amssymb,mathrsfs,url,graphicx}
\usepackage{color}

\title{Long-Time Asymptotics of a Bohmian Scalar Quantum Field in de Sitter Space-Time}

\author{
Roderich Tumulka\footnote{Department of Mathematics,
     Rutgers University, Hill Center, 
     110 Frelinghuysen Road, Piscataway, NJ 08854-8019, 
     USA. E-mail: tumulka@math.rutgers.edu}
}

\date{November 7, 2015}

\newcommand{\be}{\begin{equation}}
\newcommand{\en}{\end{equation}}
\newcommand{\dd}{\mathrm{d}}
\newcommand{\ee}{\mathrm{e}}
\newcommand{\ii}{\mathrm{i}}

\begin{document}
\maketitle
\begin{abstract}
We consider a model quantum field theory with a scalar quantum field in de Sitter space-time in a Bohmian version with a field ontology, i.e., an actual field configuration $\varphi({\bf x},t)$ guided by a wave function on the space of field configurations. We analyze the asymptotics at late times ($t\to\infty$) and provide reason to believe that for more or less any wave function and initial field configuration, every Fourier coefficient $\varphi_{\bf k}(t)$ of the field is asymptotically of the form $c_{\bf k}\sqrt{1+{\bf k}^2 \exp(-2Ht)/H^2}$, where the limiting coefficients $c_{\bf k}=\varphi_{\bf k}(\infty)$ are independent of $t$ and $H$ is the Hubble constant quantifying the expansion rate of de Sitter space-time. In particular, every field mode $\varphi_{\bf k}$ possesses a limit as $t\to\infty$ and thus ``freezes.''
This result is relevant to the question whether Boltzmann brains form in the late universe according to this theory, and supports that they do not.

\medskip

\end{abstract}

\section{Introduction}

We consider a Hermitian scalar quantum field on de Sitter space-time; the de Sitter metric reads
\be
\dd s^2 = \dd t^2 -  e^{2Ht} \delta_{ij} \dd x^i \dd x^j  \,,
\label{10}
\en
where $H$ is a constant called the Hubble constant (and we set $c=1=\hbar$). The coordinates $(t,x^1,x^2,x^3)$ cover only half of de Sitter space-time \cite{HE73}, but that is sufficient for our purposes, since we are interested in the late universe.

The quantum field theory (QFT) we use here, adopted from~\cite{PS96,GST15}, can be thought of as arising by quantization of the classical field equation
\be
{\ddot \varphi} + 3H  {\dot \varphi} - \ee^{-2Ht} \nabla^2 \varphi = 0
\label{11}
\en
for a real scalar field $\varphi({\bf x},t)$, where the dots denote $t$-derivatives. The state vector $\Psi$ can be expressed in the functional Schr\"odinger picture as a time-dependent functional $\Psi(\varphi,t)$ of the field configuration, evolving according to a Schr\"odinger equation given explicitly in \eqref{Psiflrw} below.

In Bohmian mechanics \cite{Bohm52,Gol01b}, one postulates the existence of an actual configuration obeying an equation of motion involving the wave function. Extensions of Bohmian mechanics to quantum field theory use either a particle ontology \cite{Bell86,DGTZ04,DGTZ05,CS07} or a field ontology \cite{Bohm52,Str10,Str11}; we choose here the latter. Bohm-type theories need a preferred foliation of space-time into spacelike hypersurfaces, usually assumed to be determined by a covariant law \cite{DGNSZ14}; here we use the foliation given by the hypersurfaces $t=\mathrm{const.}$. The Bohmian version of the QFT considered here is adopted from \cite{HAM95,PNSS12,GST15}; the equation of motion for the actual field configuration $\varphi({\bf x},t)$ is given in \eqref{fieldguidance} below. The field configuration $\varphi$ can be expressed either as a function $\varphi({\bf x},t)$ of the position coordinate $\bf x$ or through its Fourier transform $\varphi_{\bf k}(t)$, defined by
\begin{equation}
\varphi_{\bf k}=\int{\frac{\dd^3x}{(2\pi)^{3/2}}\, \varphi({\bf x}) \, \ee^{-\ii {\bf k}
\cdot {\bf x}}} \,.
\label{13varphi}
\end{equation}

We derive that for more or less any\footnote{It is hard to give an exact condition on the wave function and the initial field configuration under which \eqref{asymptote} is valid. But our reasoning suggests that it is very commonly valid, at least as commonly as Bohmian trajectories of non-relativistic oscillators are smooth---a property usually taken for granted in the physics literature.} wave function and any initial field configuration, the asymptotic long-time behavior of $\varphi$ is
\be\label{asymptote}
\varphi_{\bf k}(t) \approx c_{\bf k} \sqrt{1+{\bf k}^2 \exp(-2Ht)/H^2} \quad \text{for }t>t_0\,,
\en 
where $t_0$ is independent of $\bf k$ (but depends on the wave function), and
\be\label{cdef}
c_{\bf k} = \lim_{t\to\infty} \varphi_{\bf k}(t)
\en
is independent of $t$. The $\approx$ sign in \eqref{asymptote} means that the relative error is small and tends to 0 as $t\to\infty$. As a consequence, the limit \eqref{cdef} exists, which means that the field mode is ``frozen'' (i.e., time-independent) at late times. Similar results have been obtained with different methods by Ryssens \cite{Rys12}.

It is relevant that $t_0$ is independent of ${\bf k}$ because the $\bf x$ are comoving coordinates, so that (e.g.)\ a centimeter length in space corresponds to smaller and smaller differences in ${\bf x}$ as $t\to\infty$, and a centimeter wave length corresponds to larger and larger values of ${\bf k}$. If $t_0$ depended on $\bf k$, then it would not be clear if \eqref{asymptote} ever, for arbitrarily large $t$, applied on the centimeter scale.

The phenomenon of freezing at late times is connected to the expansion of de Sitter space-time and does not occur in Minkowski space-time. However, it is known already from non-relativistic Bohmian mechanics that systems whose wave function is a non-degenerate energy eigenfunction are frozen forever. The phenomenon of freezing at late times is relevant to the absence of a Boltzmann brain problem in the Bohmian approach to cosmology, as discussed in Section~\ref{sec:BB} below and more extensively in \cite{GST15}. Put briefly, it is believed that our universe will, at late times, have approximately a de Sitter geometry. Freezing at late times (or the simple motion according to \eqref{asymptote}) ensures that Boltzmann brains (i.e., brains that come into existence from particles in a random state through an extreme coincidence) are outnumbered by far by normal brains (i.e., brains that come into existence through familiar biological processes and, prior to that, evolution of life forms), and therefore that the ``Boltzmann brain problem'' (i.e., the problem that the theory might imply the wrong prediction that we are Boltzmann brains) is avoided.

It would also be of interest to extend the result of this paper to other space-time geometries besides de Sitter, in particular to other spatially flat space-times in the Friedmann-Lema\^itre-Robertson-Walker (FLRW) family, whose metric reads
\be
\dd s^2 = \dd t^2 -  a(t)^2 \, \delta_{ij} \dd x^i \dd x^j 
%= a(\eta)^2 \left(\dd \eta^2 - \delta_{ij} \dd x^i \dd x^j  \right)
\en
with some function $a(t)$ governing the expansion. On the one hand, it would then also cover the possibility that the late universe may not have a de Sitter geometry but another FLRW one; on the other hand, the result may then also apply to the early universe, which, according to inflation theory, had an FLRW geometry (but not de Sitter) during the inflationary era. It would then imply in particular that at the end of the inflationary era, the perturbations in the Bohmian fields behave classically (see \cite{PNSS12} for discussion), a desired behavior in inflation theory, and this would occur even without decoherence \cite{KP09}. In \cite{PNSS12}, this classical behavior of the Bohmian fields was confirmed for the ground state. However, at present I do not know how to generalize the result of this paper to FLRW geometries.\footnote{Here is the difficulty. It is clear that the relation \eqref{etadef} needs to be replaced by $\dd t= a\,\dd \eta$, \eqref{yHphi} by $y=a\varphi$, and factors of $1/\eta$ by $-a'/a$ in \eqref{Psiflrw}, \eqref{fieldguidance}, and \eqref{cond2}. But the present result depends on the simple explicit solution \eqref{gammadef}--\eqref{alphadef} to the coupled ordinary differential equations \eqref{cond1}--\eqref{cond3}, and with $1/\eta$ in \eqref{cond2} replaced by $-a'/a$ I do not know an analogous explicit solution.}

The remainder of this paper is organized as follows.
In Section~\ref{sec:setup}, we define the model. In Section~\ref{sec:mode-by-mode}, we consider each mode individually. In Section~\ref{sec:multi-mode}, we draw the conclusions for a general, multi-mode, non-product quantum state $\Psi$. Finally, in Section~\ref{sec:BB}, we discuss the implications of our result for the Boltzmann brain problem.

\section{Model Setup}
\label{sec:setup}

Apart from the ``cosmic time'' $t$, we also use the ``conformal time'' $\eta$ defined by
\be
\dd t=\ee^{Ht} \,\dd \eta\,,
\en 
explicitly
\be\label{etadef}
\eta=-H^{-1}\ee^{-Ht} \quad\text{or}\quad t=-H^{-1}\log(-H\eta)\,.
\en
Note that $-\infty<\eta<0$ while $-\infty<t<\infty$, and that $\eta$ is an increasing function of $t$, so that the limit $t\to\infty$ corresponds to $\eta\to 0-$ (i.e., approaching 0 from the left). In the coordinates $(\eta,x^1,x^2,x^3)$, the de Sitter metric reads
\be
\dd s^2 =  \frac{1}{H^2\eta^2} \left(\dd
\eta^2 - \delta_{ij} \dd x^i \dd x^j  \right) \,.
\label{10eta}
\en

We consider a real scalar field $\varphi({\bf x},t)$. (The expressions $\varphi({\bf x},t)$ and $\varphi({\bf x},\eta)$ will denote different functions on $\mathbb{R}^4$ representing the same field on space-time.) It is common to express the field in terms of Fourier modes $\varphi_{\bf k}$ as in \eqref{13varphi} and to use the rescaled field variable
\be\label{yHphi}
y=\ee^{Ht}\varphi\,;
\en
we have that 
$y^*_{\bf k} = y_{-{\bf k}}$ (because $y({\bf x})$ is real). 

The Schr\"odinger equation for
the wave functional $\Psi(y,\eta)$ reads
\be\label{Psiflrw}
\ii \frac{\partial\Psi}{\partial\eta} = 
\frac{1}{2} \int \dd^3x \left[ -\frac{\delta^2}{\delta y({\bf x})^2} + \delta_{ij}\, \partial^i y({\bf x})\,
\partial^j y({\bf x}) + \frac{\ii}{\eta}\left(\frac{\delta}{\delta y({\bf x})} y({\bf x}) +
y({\bf x})\frac{\delta}{\delta y({\bf x})} \right)\right]\Psi \,.
\en
Equivalenty, in terms of Fourier modes $y_{\bf k}$, the Schr\"odinger equation
becomes
\begin{equation}\label{Psiflrwk}
\ii \frac{\partial\Psi}{\partial\eta} = 
\int_{{\mathbb R}^{3+}} \dd^3k \left[ -\frac{\delta^2}{\delta y_{\bf k}^*\delta
y_{\bf k}}+
k^2 \, y_{\bf k}^* \, y_{\bf k}
+ \frac{\ii}{\eta}\left(\frac{\delta}{\delta y_{\bf k}^*}y_{\bf k}^*+
y_{\bf k}\frac{\delta}{\delta y_{\bf k}}\right)\right]\Psi\,,
\end{equation}
where $k=|{\bf k}|$, and the variational derivatives $\delta/\delta y_{\bf k}$ and $\delta/\delta y_{\bf k}^*$ are to be understood in the sense of Wirtinger derivatives (i.e., for a complex variable $z=x+iy$, $\partial_z = \tfrac{1}{2}\partial_x -\tfrac{\ii}{2} \partial_y$ and $\partial_{z^*}=\tfrac{1}{2}\partial_x + \tfrac{\ii}{2} \partial_y$). Since $y_{-\bf k}=y_{\bf k}^*$, we restrict the integration to half the number of possible
modes (denoted by ${\mathbb R}^{3+}$), so that only independent ones are
included. 

The Bohmian version of the QFT considered here involves an actual scalar field $\varphi({\bf x},t)$ on de Sitter space-time that can be equivalently expressed as $y({\bf x},\eta)$ or $y_{\bf k}(\eta)$ (corresponding to different coordinates in the space of field configurations) and obeys the equation of motion
\be\label{fieldguidance}
y({\bf x})' = \frac{\delta S}{\delta y ({\bf x})} -  \frac{1}{\eta}y({\bf x})
\en
or, in terms of Fourier modes,
\be\label{fieldguidancek}
y'_{\bf k}= \frac{\delta S}{\delta y^*_{\bf k}}-\frac{1}{\eta}y_{\bf k}\,,
\en
where the prime means $\eta$-derivative and $S$ the phase of the wave function, $\Psi = |\Psi| \ee^{\ii S}$. This equation of motion, adopted from \cite{HAM95,PNSS12}, can be obtained via certain general rules \cite{SV} for obtaining a Bohm-type equation of motion from a given Hamiltonian. It entails (on a formal, non-rigorous level) that the $|\Psi|^2$ density is preserved, as usual in Bohm-type theories \cite{Gol01b}.

\section{Mode-By-Mode Investigation}
\label{sec:mode-by-mode}

Since the Hamiltonian in \eqref{Psiflrw} is an integral of terms involving only one ${\bf k}$, different modes do not interact. Thus, if (as we assume in this section) the initial wave functional is of the product form
\be
\Psi = \prod_{{\bf k} \in {\mathbb R}^{3+}} \Psi_{\bf k}(y_{\bf k},\eta) \,, 
\label{50}
\en
then it will be a product for all times, and the functional Schr\"odinger equation reduces to the following partial
differential equation for each $\Psi_{\bf k}$:
\begin{equation}\label{Psiketa}
\ii\frac{\partial\Psi_{\bf k}}{\partial\eta}=
\left[ -\frac{\partial^2}{\partial y_{\bf k}^*\partial y_{\bf k}}+
k^2 y_{\bf k}^* y_{\bf k}
+\frac{\ii}{\eta}\left(\frac{\partial}{\partial y_{\bf k}^*}y_{\bf
k}^*+
y_{\bf k}\frac{\partial}{\partial y_{\bf k}}\right)\right]\Psi_{\bf k}\,.
\end{equation}
The equation of motion reduces to
\be\label{reducedfieldguidance}
y'_{\bf k}= \frac{\partial S_{\bf k}}{\partial y^*_{\bf k}}-\frac{1}{\eta}y_{\bf k}
\,,
\en
where $\Psi_{\bf k} = |\Psi_{\bf k}| \ee^{\ii S_{\bf k}}$. In particular, the motion of $y_{\bf k}$ is independent of the other $y_{\bf k'}$ and the other $\Psi_{\bf k'}$.

Let us consider a single factor $\Psi_{\bf k}(y_{\bf k},\eta)$, for simplicity now denoted $\Psi(y,\eta)$, in the limit as $\eta$ approaches 0 from the left, $\eta \to 0-$ (which corresponds to $t\to\infty$), and get control of the asymptotics of solutions of \eqref{Psiketa}.
We introduce the rescaled field variable
\be
z=\gamma(\eta)^{-1} y
\en
and the rescaled and phase-transformed wave function
\be
\Phi(z,\eta) = \ee^{\alpha(\eta)}\, \ee^{\ii \beta(\eta) z^*z} \, \Psi\bigl(\gamma(\eta)\, z,\eta\bigr)\,,
\en
where the real-valued functions $\alpha(\eta),\beta(\eta),\gamma(\eta)$ will be chosen later. Computing $\ii\partial \Phi/\partial \eta$, we find that
\be\label{Phi1}
\ii\frac{\partial \Phi}{\partial \eta} = -\frac{1}{\gamma^2} \frac{\partial^2 \Phi}{\partial z^* \partial z}
+\biggl( -\beta'+k^2\gamma^2 - \frac{\beta^2}{\gamma^2} \biggr) z^*z \,\Phi
\en
plus two further terms with prefactors given by the left-hand sides of the following equations:
\begin{align}
\alpha' 
%+ \frac{\beta}{\gamma^2} + \frac{1}{\eta} 
-\frac{\gamma'}{\gamma} &=0 \label{cond1}\\
\frac{\gamma'}{\gamma} + \frac{\beta}{\gamma^2} + \frac{1}{\eta} &=0\,. \label{cond2}
\end{align}
If we demand \eqref{cond1} and \eqref{cond2} and, in addition,
\be
-\beta' +k^2\gamma^2- \frac{\beta^2}{\gamma^2}  = \frac{\omega^2}{\gamma^2}\,, \label{cond3}
\en
with some constant $\omega\geq 0$ to be chosen later, that is, if we choose $\alpha,\beta,\gamma$ to be a (real-valued) solution of the system \eqref{cond1}--\eqref{cond3} of ordinary differential equations with $\gamma>0$, then $\Phi$ evolves according to
\be\label{Phi2}
\ii\frac{\partial \Phi}{\partial \eta} = -\frac{1}{\gamma^2} \frac{\partial^2 \Phi}{\partial z^* \partial z}
+\frac{\omega^2}{\gamma^2}  z^*z \, \Phi\,.
\en
In terms of a rescaled time variable $\tau$ defined by
\be\label{taudef}
\dd\tau = \gamma(\eta)^{-2}\, \dd\eta
\en
with origin $\tau=0$ chosen as corresponding to $\eta=0$, the evolution equation \eqref{Phi2} becomes
\be\label{Phitau}
\ii\frac{\partial \Phi}{\partial \tau} = -  \frac{\partial^2 \Phi}{\partial z^* \partial z} + \omega^2 z^*z \,\Phi\,,
\en
i.e., the non-relativistic Schr\"odinger equation in a harmonic oscillator potential, with the configuration space given by the complex plane and $m=2$ (note that $\partial_{z^*} \partial_z = \tfrac{1}{4}(\partial_x^2+\partial_y^2)$). For the field variable 
\be\label{zvarphi}
z(\tau)=\gamma(\tau)^{-1} y_{\bf k}(\tau) = -\frac{1}{H\,\eta(\tau)\,\gamma(\tau)}\, \varphi_{\bf k}(\tau)\,,
\en
the equation of motion \eqref{reducedfieldguidance} becomes
\be\label{ztau}
\frac{\dd z}{\dd \tau} =  \frac{\partial}{\partial z^*} \mathrm{Im} \log \Phi(z,\tau)\,,
\en
which is equivalent to the equation of motion of non-relativistic Bohmian mechanics in the complex plane. Since the Schr\"odinger equation \eqref{Phitau} does not become singular at $\tau=0$, and the equation of motion does not either, $z(\tau)$ possesses a limit as $\tau\to 0-$, and thus as $\eta\to0-$ or $t\to\infty$. (In fact, the solutions $\Phi(\tau)$ and $z(\tau)$ to \eqref{Phitau} and \eqref{ztau} can be extended to $\tau>0$.)
That is the core of the reasoning.

If (as it will turn out to be the case) also $\eta\gamma$ possesses a limit as $\eta \to 0-$, then, by \eqref{zvarphi}, also $\varphi_{\bf k}$ possesses a limit as $\eta\to0-$ or $t\to\infty$.  
Since the limit of $z(\tau)$ as $\tau\to0-$ is typically non-zero (and that of $\eta\gamma$, too), also the limit of $\varphi_{\bf k}$ is typically non-zero. 

To obtain the asymptote \eqref{asymptote}, we consider 
the following particular solution $\alpha,\beta,\gamma$ of \eqref{cond1}--\eqref{cond3}. 
For $-\infty<\eta<0$, let
\begin{align}
\omega &= k \\
\gamma(\eta) &= -\frac{ \sqrt{1+k^2\eta^2} }{k \eta} \label{gammadef} \\
\beta(\eta) &= - \frac{1}{\eta} \label{betadef} \\
\alpha(\eta) &= \log \gamma(\eta) +\alpha_0 \label{alphadef}\\
%-\int_{\eta_i}^\eta \Bigl(\frac{\beta(\tilde\eta)}{\gamma(\tilde\eta)^2} +\frac{1}{\tilde\eta} \Bigr)\, \dd\tilde\eta\\
\tau(\eta) &= \eta-\frac{1}{k} \arctan (k\eta) \label{taueta}
\end{align} 
with %arbitrary $\eta_i<0$. 
suitable normalizing constant $\alpha_0$.
One easily verifies that \eqref{cond1}--\eqref{cond3} and \eqref{taudef} are satisfied and that $\gamma>0$ and $\beta>0$.\footnote{The Bunch-Davies state (see, e.g., Eq.s (25) and (27) in \cite{GST15}) corresponds to $\Phi$ being the ground state of \eqref{Phitau}.} 

Let $0<\varepsilon\ll 1$ be the relative error we want to allow. Since $z(0)=\lim_{\tau\to0-}z(\tau)$, there is $\tau_0<0$ such that $|z(\tau)-z(0)|<\varepsilon z(0)$ for all $\tau$ between $\tau_0$ and $0$. 
Since $\Phi(z,\tau)$ is non-singular (nothing special happens in \eqref{Phitau} and \eqref{ztau} at $\tau=0$), both $z(\tau)$ and $\dd z/\dd\tau$ are of order 1 for $-1<\tau\leq 0$; we suppose, for simplicity and definiteness, that $\tau_0$ can be chosen independently of $\bf k$. This is a rather mild assumption on $\Psi$; it follows, for example, if there is a positive constant $C$ independent of $\bf k$ such that
\be\label{zassumption}
\biggl| \frac{\dd z}{\dd\tau} \biggr| <C
\en
for $-1<\tau<0$.

Set $\eta_0=\tau_0$ and consider the interval $\eta_0<\eta<0$; since $0<\dd\tau/\dd\eta=\gamma^{-2}=1-1/(1+k^2\eta^2)<1$ independently of $\bf k$, we have that $\tau_0<\tau(\eta)<0$ for any $\eta$ between $\eta_0$ and $0$. Therefore, using \eqref{zvarphi} and $z(0)=k\varphi_{\bf k}(0)/H = kc_{\bf k}/H$,
%\be
%\Bigl|-\frac{\varphi_{\bf k}(\eta)}{H\eta \gamma}-\frac{k\,c_{\bf k}}{H}\Bigr|<\varepsilon c_{\bf k} k/H\,
%\en
\be\label{estimate}
\Biggl|\frac{\varphi_{\bf k}(\eta)-c_{\bf k}\sqrt{1+k^2\eta^2}}{c_{\bf k}\sqrt{1+k^2\eta^2}}\Biggr|<\varepsilon
\en
for any $\eta$ between $\eta_0$ and $0$, which is the desired relation \eqref{asymptote}.

\section{General Multi-Mode Wave Function}
\label{sec:multi-mode}

Returning from $\Psi=\Psi_{\bf k}$ to a general wave functional $\Psi$ that depends on all $y_{\bf k}$ and is not of product form, an analogous reasoning goes through with
\be
z_{\bf k} = \gamma_{\bf k}(\eta)^{-1} \, y_{\bf k}
\en
(writing $\gamma_{\bf k}$ etc.\ for $\gamma$ etc.\ as in \eqref{gammadef} etc.)\ and
\be
\Phi(\eta) = \ee^{\int \dd^3k\, (\alpha_{\bf k}(\eta) +\ii \beta_{\bf k}(\eta) z_{\bf k}^* z_{\bf k})} \, \Psi(\eta)
\en
with $\int \dd^3k = \int_{\mathbb{R}^{3+}}\dd^3k$ and leads to
\begin{align}
\label{Phiketa}
\ii\frac{\partial \Phi}{\partial \eta} &= \int \frac{\dd^3k}{\gamma_{\bf k}^2} \biggl[- \frac{\delta^2}{\delta z_{\bf k}^* \delta z_{\bf k}}
+  \omega_{\bf k}\,z_{\bf k}^*z_{\bf k}\biggr] \Phi\\
\label{zketa}
\frac{\dd z_{\bf k}}{\dd \eta} &= \frac{1}{\gamma_{\bf k}^2} \frac{\delta}{\delta z_{\bf k}^*} \mathrm{Im} \log \Phi(\eta)\,.
\end{align}
The rescaled time coordinate $\tau$ introduced in \eqref{taudef} depended on $\bf k$, so this rescaling cannot be done straightforwardly in this case that involves various $\bf k$ values. Still, after solving \eqref{Phiketa} and, for all $\bf k$ simultaneously, \eqref{zketa}, thus obtaining $\Phi(\eta)$ and $z_{\bf k}(\eta)$ for all $\bf k$, we can consider $\dd z_{\bf k}/\dd\tau_{\bf k}$ (with $\dd \tau_{\bf k}= \gamma_{\bf k}^{-2} \dd\eta$), and we would like to argue that it is of order 1. To this end, let $U_{\bf k}(\tau_2-\tau_1)$ be the unitary time evolution operator solving \eqref{Phitau} with $\omega=\omega_{\bf k}=k$ from time $\tau_1$ to time $\tau_2$. Then the unitary time evolution operator solving \eqref{Phiketa} from time $\eta_1$ to time $\eta_2$ is
\be
U(\eta_1,\eta_2) = \bigotimes_{{\bf k}\in \mathbb{R}^{3+}} U_{\bf k} \bigl(\tau_{\bf k}(\eta_2)-\tau_{\bf k}(\eta_2)\bigr)\,.
\en
This operator does not become singular as $\eta_2\to 0-$ (in fact, it can even be extended smoothly to positive $\eta_2$) because $\tau_{\bf k}(0)=0$ (in fact, the definition of $\tau_{\bf k}$ in \eqref{taueta} extends smoothly to positive $\eta$) and $U_{\bf k}$ is non-singular for any $\tau_2-\tau_1$, due to the non-singular nature of \eqref{Phitau}.  Therefore, $\lim_{\eta\to0-}\Phi(\eta)$ exists for the solution $\Phi(\eta)$ of \eqref{Phiketa} (in fact, $\Phi(\eta)$ can be extended smoothly to positive $\eta$). Thus, $(\delta/\delta z_{\bf k}^*) \mathrm{Im} \log \Phi(\eta)$, which equals $\dd z_{\bf k}/\dd\tau_{\bf k}$, should not blow up as $\eta\to0-$, it should indeed be of order 1. For simplicity and definiteness, we assume that there is $\tau_0<0$ such that $|z_{\bf k}(\tau_{\bf k})-z_{\bf k}(0)|<\varepsilon z_{\bf k}(0)$ for all $\bf k$ and all $\tau_{\bf k}$ between $\tau_0$ and $0$. This follows, for example, if there is a positive constant $C$ such that
\be\label{zassumption2}
\Bigl| \frac{\delta}{\delta z_{\bf k}^*} \mathrm{Im} \log \Phi(\eta) \Bigr| <C
\en
for all $\bf k$ and all $\eta$ between $-1$ and 0.

As before, we set $\eta_0=\tau_0$ and obtain that for $\eta$ between $\eta_0$ and 0, \eqref{estimate} and thus \eqref{asymptote} holds, which is what we claimed.

\section{Implications for the Boltzmann Brain Problem}
\label{sec:BB}

In a classical gas in a box (of the size of a building, say) in thermal equilibrium, the particles may through extreme coincidence happen to come together in such a way as to form a functioning human brain, called a \emph{Boltzmann brain} \cite{AS04,BCP15,GST15}, for example in the same macrostate as your brain right now. Such a Boltzmann brain will have the same memories as you, but they will be false memories, as the events remembered never happened to that brain. The formation of a Boltzmann brain is extremely unlikely to occur within an hour (or even a century) in our box, but if we wait for a very long time, much longer than the present age of the universe, then it is likely to happen at some point in time. The larger the box, the more frequently it will occur. This event is a special case of a fluctuation out of thermal equilibrium in which entropy happens to decrease, contrary to the overwhelmingly likely tendency of entropy to increase. Boltzmann brains die quickly, most of them for lack of oxygen, but for a short while they may function.

This theoretical scenario becomes relevant if our universe continues to exist as $t\to\infty$, reaches universal thermal equilibrium at some point in the distant future, and remains there except for fluctuations. Then it is overwhelmingly likely that Boltzmann brains will arise, in fact over and over. So do ``Boltzmann planets'' and ``Boltzmann galaxies,'' but those are not by themselves problematical. A problem arises from the existence of Boltzmann \emph{brains}, specifically from the fact that, if the universe continues to exist forever (or only long enough), very many Boltzmann brains will form in the long run, much more than the total number of ``normal'' brains that ever existed, where normal brains are those which came into existence as parts of human beings that had parents and ancestors and arose from an evolution of life forms. For example, suppose that the spatial volume of the universe is finite, that the universe began at some time $t_0$, and that the entire universe reaches thermal equilibrium at some time $T$; then the total number of normal brains is finite, while the number of Boltzmann brains grows to infinity as $t\to\infty$. Or so would be the prediction of the physical theory we are using here. The problem with that is that this physical theory would thus also predict that \emph{we} are Boltzmann brains. After all, the \emph{Copernican principle} \cite{Gott} asserts that the correct way of extracting predictions from a physical theory is that we should see what typical observers see. But we are sure that we are not Boltzmann brains, certainly more sure than we are of any given physical theory; for example, we trust our memories a long way; for another example, the overwhelming majority of Boltzmann brains find themselves surrounded by thermal equilibrium, and we do not, so we have empirical evidence that we are not Boltzmann brains. The question is, how can any serious theory avoid predicting that we should be Boltzmann brains?

Various models of cosmology predict that the late universe will be approximately a de Sitter space-time, and will in particular continue to exist forever. The quantum state of the matter fields at late times is often taken to be the Bunch-Davies state. If that is so, will the late universe contain a large number of Boltzmann brains or not? That depends on the interpretation of quantum theory. In the Bohmian version of quantum theory, if, as suggested by the results of this paper, ``freezing'' occurs at late times, then Boltzmann brains do not arise because, even if a brain configuration occurs somewhere, the brain will not be functional because it is frozen. (In addition, for most $\Psi$ including the Bunch-Davies state, the $|\Psi|^2$ probability for a brain configuration to occur at all within our Hubble volume is tiny.) This shows how it may be that the Boltzmann brain problem does not arise in the Bohmian approach. Further discussion within the Bohmian approach can be found in \cite{GST15}, and within the Everettian approach in \cite{BCP15}.

\bigskip

\noindent\textit{Acknowledgments.} 
The author acknowledges support from the John Templeton Foundation, grant no.\ 37433. I thank Shelly Goldstein and Ward Struyve for helpful discussions and two anonymous referees for their suggestions.


\begin{thebibliography}{10}

\bibitem{AS04} A. Albrecht and L. Sorbo:
	Can the universe afford inflation?
	\textit{Physical Review D} \textbf{70}: 063528 (2004)
	\url{http://arxiv.org/abs/hep-th/0405270}

\bibitem{Bell86} J. S. Bell:
	Beables for quantum field theory.
	\textit{Physics Reports} \textbf{137}: 49--54 (1986).  
	Reprinted on p.~173 in J. S. Bell: \textit{Speakable and unspeakable in quantum mechanics,}
	Cambridge University Press (1987)  
%	Also reprinted on p.~227 in F. D. Peat and B. J. Hiley (eds):
%	\textit{Quantum Implications: Essays in Honour of David Bohm.} 
%	London: Routledge (1987).
%	Also reprinted as chap.~17 in M. Bell, K. Gottfried, and M. Veltman (eds):
%	\textit{John S.\ Bell on the Foundations of Quantum Mechanics.}
%	World Scientific Publishing (2001).

\bibitem{BCP15} K.K. Boddy, S.M. Carroll, and J. Pollack:
	Why Boltzmann Brains DonÕt Fluctuate Into Existence From the De Sitter Vacuum.
	In K. Chamcham, J. Barrow, J. Silk, and S. Saunders (editors), 
	{\it The Philosophy of Cosmology}, Cambridge University Press (2016)
	\url{http://arxiv.org/abs/1505.02780}

\bibitem{Bohm52} D. Bohm: 
	A Suggested Interpretation of the Quantum Theory in Terms of ``Hidden'' Variables, I and II. 
	\textit{Physical Review} \textbf{85}: 166--193 (1952)

\bibitem{CS07} S. Colin and W. Struyve:
	A Dirac sea pilot-wave model for quantum field theory.
	{\it Journal of Physics A: Mathematical and Theoretical} {\bf 40}: 7309-7342 (2007)
	\url{http://arxiv.org/abs/quant-ph/0701085}

\bibitem{DGNSZ14} D. D\"urr, S. Goldstein, T. Norsen, W. Struyve, and N. Zangh\`\i:
	Can Bohmian mechanics be made relativistic?
	\textit{Proceedings of the Royal Society A} \textbf{470}: 20130699 (2014)
	\url{http://arxiv.org/abs/1307.1714}

\bibitem{DGTZ04} D. D\"urr, S. Goldstein, R. Tumulka, and N. Zangh\`\i:
	Bohmian Mechanics and Quantum Field Theory.
	\textit{Physical Review Letters} \textbf{93}: 090402 (2004)
	\url{http://arxiv.org/abs/quant-ph/0303156}

\bibitem{DGTZ05} D. D\"urr, S. Goldstein, R. Tumulka, and N. Zangh\`\i:
	Bell-type quantum field theories.
	\textit{Journal of Physics A: Mathematical and General} \textbf{38}: R1--R43 (2005)
	\url{http:arxiv.org/abs/quant-ph/0407116}

\bibitem{Gol01b} S. Goldstein: 
	Bohmian Mechanics. 
	In E. N. Zalta (ed.), \textit{Stanford Encyclopedia of Philosophy}, 
	published online by Stanford University (2001) 
	\url{http://plato.stanford.edu/entries/qm-bohm}

\bibitem{Gott} J.R. Gott:
	Implications of the Copernican principle for our future prospects.
	\textit{Nature} \textbf{363}: 315-319 (1993)

\bibitem{GST15} S. Goldstein, W. Struyve, and R. Tumulka:
	The Bohmian Approach to the Problems of Cosmological Quantum Fluctuations.
	To appear in A. Ijjas and B. Loewer (editors): 
	\textit{Guide to the Philosophy of Cosmology},
	Oxford University Press (2016)
	\url{http://arxiv.org/abs/1508.01017}

\bibitem{HE73} S.W. Hawking and G.F.R. Ellis:
	\textit{The large-scale structure of space-time.}
	Cambridge University Press (1973)

\bibitem{HAM95} B.J. Hiley and A.H. Aziz Mufti:
	The Ontological Interpretation of Quantum Field Theory Applied in a Cosmological Contex.
	Pages 141--156 in M. Ferrero and A. van der Merwe (editors): 
	\textit{Fundamental Theories of Physics} \textbf{73},
	Dordrecht: Kluwer (1995)

\bibitem{KP09} C. Kiefer and D. Polarski:
	Why do cosmological perturbations look classical to us?
	\textit{Advanced Science Letters} \textbf{2}: 164--173 (2009)
	\url{http://arxiv.org/abs/0810.0087}

\bibitem{PNSS12} N. Pinto-Neto, G. Santos, and W. Struyve:
	The quantum-to-classical transition of primordial cosmological perturbations.
	\textit{Physical Review D} \textbf{85}: 083506 (2012)
	\url{http://arxiv.org/abs/1110.1339}

\bibitem{PS96} D. Polarski and A.A. Starobinsky:
	Semiclassicality and Decoherence of Cosmological Perturbations.
	\textit{Classical and Quantum Gravity} \textbf{13}: 377--392 (1996) 
	\url{http://arxiv.org/abs/gr-qc/9504030v2}

\bibitem{Rys12} W. Ryssens:
	{\it On the Quantum-to-Classical Transition of Primordial Perturbations.}
	Master thesis, Department of Physics and Astronomy, Katholieke Universiteit Leuven (2012)

\bibitem{Str10} W. Struyve:
	Pilot-wave theory and quantum fields.
	\textit{Reports on Progress in Physics} \textbf{73}: 106001 (2010)
	\url{http://arxiv.org/abs/0707.3685}

\bibitem{Str11} W. Struyve:
	Pilot-wave approaches to quantum field theory.
	\textit{Journal of Physics: Conference Series} \textbf{306}: 012047 (2011)
	\url{http://arxiv.org/abs/1101.5819}

\bibitem{SV} W. Struyve and A. Valentini:
	De Broglie-Bohm Guidance Equations for Arbitrary Hamiltonians.
	\textit{Journal of Physics A: Mathematical and Theoretical} \textbf{42}: 035301 (2009)
	\url{http://arxiv.org/abs/0808.0290}

\end{thebibliography}
\end{document}